\documentstyle[graphicx,aps,pre,twocolumn]{revtex}

\newcommand{\myscalebox}[1]{\scalebox{0.4}[0.45]{#1}}
\begin{document}
\title{Calculation of ground states of four-dimensional $\pm J$ 
Ising spin glasses}

\author{Alexander K. Hartmann\\
{\small  hartmann@theorie.physik.uni-goettingen.de}\\
{\small Institut f\"ur theoretische Physik, Bunsenstr. 9}\\
{\small 37073 G\"ottingen, Germany}\\
{\small Tel. +49-551-395995, Fax. +49-551-399631}}

\date{\today}
\maketitle
\begin{abstract}
Ground states of four-dimensional (d=4) EA Ising spin glasses are calculated
for sizes up to $7 \times 7 \times 7 \times 7$ using a combination 
of a genetic algorithm and cluster-exact
approximation. The ground-state energy of the infinite system is
extrapolated as $e_0^{\infty}=-2.095(1)$.
The ground-state stiffness (or domain wall) energy $\Delta$ is calculated. A
$|\Delta| \sim L^{\Theta_S}$ behavior with $\Theta_S=0.65(4)$ is found which
confirms that the d=4 model has an equilibrium 
spin-glass-paramagnet transition for non-zero $T_c$.

{\bf Keywords (PACS-codes)}: Spin glasses and other random models (75.10.Nr), 
Numerical simulation studies (75.40.Mg),
General mathematical systems (02.10.Jf). 
\end{abstract}


\section{Introduction}

Optimization methods have found widespread application in
computational physics. Among these the investigation of the
low-temperature behavior of spin glasses  \cite{binder86}
attracted most of the attention within the statistical physics community.
The reason is that despite its simple definition (see below) its
behavior is far from being understood. From the computational point of
view the calculation of spin-glass ground states is very demanding,
because it belongs to the class of the NP-hard problems
\cite{barahona82}. This means that only algorithms are available,
for which the running time on a computer  increases
exponentially with the system size. In this work a method recently
proposed, the {\em cluster-exact approximation} (CEA) \cite{alex2} is
applied to four-dimensional Ising spin glasses.

The model under investigation here consists of $N$ spins 
$\sigma_i = \pm 1$, described by the Hamiltonian
\begin{equation}
H \equiv - \sum_{\langle i,j\rangle} J_{ij} \sigma_i \sigma_j
\end{equation}
where $\langle \ldots \rangle $ denotes a sum over pair of nearest neighbors.
In this report simple 4d lattices  are considered, i.e. 
$N=L^4$.  The nearest neighbor interactions (bonds) take
independently $J_{ij} = \pm 1$ with equal probability. 
Periodic boundary conditions are applied to the systems. No kind of 
external magnetic field is present here.

Four-dimensional Ising spin glasses have been investigated rather
rarely. Most of the results were obtained via Monte-Carlo (MC)
simulations at finite temperature, see e.g. 
\cite{bhatt85u88,reger90,badoni93,parisi96,bernardi97,marinari98,hukushima99}.
Here the $T=0$ behavior is investigated, i.e. ground states are
calculated. This has the advantage, that one does not
encounter ergodicity problems or critical
slowing down like in algorithms which base on MC methods. Only
one attempt \cite{wanschura96} to address the 4d 
spin-glass ground-state problem is known to the author. 
But, as we will see later, the former results
suffer from the problem, that not the true global minima of the 
energy were obtained. Furthermore, no analytic predictions of
the ground-state energy have been noted by the author.

The question whether finite-dimensional  Ising spin glasses 
show an ordered phase below
a non-zero transition temperature $T_c$ is of crucial interest. By MC
simulations around the (expected) transition temperature this question
is hard to solve.
Another way to  address this question is to calculate the {\em stiffness} or
{\em domain wall energy} $\Delta=E^a-E^p$ which is the difference 
between the ground-state energies $E^a, E^p$ 
for antiperiodic and periodic boundary conditions
in one direction\cite{bray84,mcmillan84}. Here the antiperiodic 
boundary conditions for calculating $E^a$ are realized by inverting one
plane of bonds. For the other
directions periodic boundary conditions are applied always. This treatment
introduces a domain wall into the system. If a model exhibits an
ordered low-temperature phase, the domain wall increases with growing system
size, which becomes visible through the behavior of $\Delta$:
the disorder-averaged stiffness energy shows a finite size dependence
\begin{equation}
\langle |\Delta| \rangle \sim L^{\Theta_S}
\end{equation}
A positive value of the stiffness exponent $\Theta_S$ indicates the
existence of an ordered phase for non-zero temperature. For example
a simple $d=2$ Ising ferromagnet has $\Theta_S=1$. For spin glasses, the
stiffness exponent plays additionally 
an important role within the droplet-scaling
theory \cite{mcmillan,bray,fisher86,fisher88,bovier}, 
 where it describes the finite-size behavior of the basic excitations (the
 droplets). 

Using this kind of analysis is was proven that the 2d spin glass 
exhibits no ordering for $T>0$ \cite{kawashima97}. For the
three-dimensional problem in a recent calculation \cite{alex-stiff}
 by applying genetic CEA a value of $\Theta_S=0.19(2)$ was found, 
which shows, that indeed the $d=3$ model
has a spin-glass phase for nonzero temperature. For $d=4$ the
existence of a finite $T_c\approx 2.1$
was proven rather early even by MC simulations
\cite{bhatt85u88,reger90}, but the value for the
stiffness-exponent $\Theta_S$ is of interest on its own. In \cite{hukushima99}
recently a value of $\Theta_S=0.82(6)$ was found by performing a MC
simulation near $T_c$. In the work presented here the value is obtained via
ground-state calculations.

The paper is organized as follows: In the next section the algorithm
applied here is briefly presented. The main section contains the results
for the ground-state energy and the stiffness exponent. 
Finally a summary is given.

\section{Algorithm}

The technique for the calculation bases on a special genetic
algorithm \cite{pal96,michal92} and on cluster-exact approximation  
\cite{alex2} which is  an optimization method designed especially
for spin glasses. Now a brief description of the method is given.

Genetic algorithms are biologically motivated. An optimal
solution is found by treating many instances of the problem in
parallel, keeping only better instances and replacing bad ones by new
ones (survival of the fittest).
The genetic algorithm starts with an initial population of $M_i$
randomly initialized spin configurations (= {\em individuals}),
which are linearly arranged in
a ring. Then $\nu \times M_i$ times two neighbors from the population
are taken (called {\em parents}) and two offspring are created
using the so called triadic crossover \cite{pal95}. 
Then a mutation with a rate of $p_m$
is applied to each offspring, i.e. a fraction $p_m$ of the
spins is reversed.

Next for both offspring the energy is reduced by applying
CEA. The algorithm bases on the concept of  {\em frustration}
\cite{toulouse77}. 
The method constructs iteratively and randomly 
a non-frustrated cluster of spins, whereas
spins with many unsatisfied bonds are more likely to be added to the
cluster. 
The  non-cluster spins act like local magnetic fields on the cluster spins.
For the spins of the cluster an energetic minimum state can be 
calculated in polynomial time
by using graph-theoretical methods 
\cite{claibo,knoedel,swamy}: an equivalent network is constructed
\cite{picard1}, the maximum flow is calculated 
\cite{traeff,tarjan} and the spins of the
cluster are set to the orientations leading to a minimum in energy. 
This minimization step
is performed $n_{\min}$ times for each offspring.

Afterwards each offspring is compared with one of its parents. The
pairs are chosen in the way that the sum of the phenotypic differences
between them is minimal. The phenotypic difference is defined here as the
number of spin  where the two configurations differ. Each
parent is replaced if its energy is not lower (i.e. better) than the 
corresponding offspring.

After this creation of offspring is performed
 $\nu \times M_i$ times the population
is halved: From each pair of neighbors the configuration 
 which has the higher energy is eliminated. If not more than 4
individuals remain the process is stopped and the best individual
is taken as result of the calculation.

The whole algorithm is performed $n_R$ times and all configurations
which exhibit the lowest energy are stored, resulting in $n_g$ statistical
independent ground state configurations. The method was already applied for
the investigation of the ground-state landscape of 3d Ising spin
glasses \cite{alex-3d}.

\section{Results}

In this section at first the values for the 
simulation parameters, which are defined above, are presented. Then the
finite-size behavior of the ground-state energy is
investigated. Finally results for the stiffness energy are discussed.

The  simulation parameters were determined in the following way:
For the system sizes $L=2,4,6,7$  several different combinations
of the 
parameters $M_i, \nu, n_{min}, p_m$  were tested. 
For the final parameter sets it is not possible to obtain lower
energies even by using parameters where the calculation consumes  
four times the computational effort. For $L=3,5$ the parameter sets
for $L+1$ were used.
Using parameter sets chosen this way genetic CEA calculates true
ground states, as shown in  \cite{alex-stiff}. 
It  should be pointed out that it is relatively easy
to obtain states, which exhibit an energy slightly above the true
ground state energy. The hard task is to obtain really the global
minimum of the energy. 

Here $p_m=0.1$ and $n_R=5$ were used for all system sizes. 
Table \ref{tab_parameters} summarizes the parameters. Also the
typical computer time $\tau$ per ground state
computation on a 80 MHz PPC601 is given. 

Ground states were calculated for system sizes up to $7\times 7\times
7 \times 7$ for $N_L$
independent realizations (see table \ref{tab_parameters})
of the random variables. For each realization
the ground states with periodic and antiperiodic boundary condition
in one direction
were calculated. The remaining three directions are always subjected to
periodic boundary conditions. One can extract from the table that for
small system sizes $L\le 4$ ground states are rather easily to obtain,
while the $L=7$ systems alone required 6560 CPU-days. 
Using these parameters on average $n_g>2.7$
ground states were obtained for every system size $L$ using  $n_R=5$ runs per
realization.

The average ground-state energy $e_0$ per spin is shown in
Fig.\ref{figEnergy} as a function of the system size $L$. Using a fit
to $e_0(L)=e_0^{\infty} + a*L^{-b}$ the value for the
infinite system is extrapolated, resulting in $e_0^{\infty} = -2.095(1)$
($a=7.1(7),b=-4.2(1)$). 
This value is compatible with the lower bound of
$e_0=\sqrt{2d\ln 2}\approx 2.35$ given by the random energy model
\cite{derrida81}. The value calculated here is substantially smaller than 
the result $e_0^{\infty} =
-2.054(3)$, which was obtained in \cite{wanschura96} using a pure
genetic algorithm. This shows that in \cite{wanschura96} not the true global
minima were found, which can be concluded also from the fact, that
there $e_0(L)$ increases with growing system size. Because the
periodic boundary conditions impose additional constraints on the
systems, the opposite behavior is expected, as found for the results
presented here. For further comparison additionally some
calculations were performed by the author by 
simply rapidly quenching from random chosen spin 
configurations. By executing an analogous fit, a value of
$e_0^{\infty} = -2.04(2)$ is obtained. This shows, that the result from
\cite{wanschura96} seems to be only slightly better than the data obtained
by applying a very simple minimization method.

The distribution of the stiffness energy, which is obtained 
from performing ground-state
calculations for systems with either periodic or antiperiodic boundary
conditions in one direction, are shown in Fig. \ref{figPStiff} for
$L=5$ and $L=7$. With increasing system size the distribution
broadens. This means that larger domain walls become
more and more likely. To study this effect more quantitatively,
in Fig. \ref{figStiffness} the disorder-averaged absolute value
$\langle |\Delta| \rangle$ of the stiffness
energy is plotted as a function of the system size $L$.
Also shown is a fit $\langle|\Delta(L)|\rangle \sim L^{\Theta_S}$ 
which results in 
$\Theta_S=0.64(5)$. Here, the system sizes $L=2,3$ were left out of the
analysis, since they are below the scaling regime. 
Because of the large sample sizes the error bars are small
enough, so we can be pretty sure that $\Theta_S>0$. It confirms
earlier results from MC simulations \cite{bhatt85u88,reger90} that the 4d
EA spin glass exhibits a non-zero transition temperature $T_c$. The
value $\Theta_S=0.64(5)$ is comparable to a recent result from MC simulations
$\Theta_S=0.82(6)$ \cite{hukushima99}, 
given the facts that the system sizes are rather
small and the other result was obtained at finite temperature near the
transition point $T_c\approx 2.1$. Additionally, the prediction from
droplet-scaling theory $\Theta_S < (d-1)/2 = 1.5$ \cite{fisher88} is
fulfilled. 

It should be pointed out, that the method described above does not
guaranty to find exact ground states, although the method for choosing
the parameters makes it very likely. If states with a slightly higher
energy are obtained, the result for $e_0^{\infty}$ is not affected
very much. For the stiffness energy, it was shown in \cite{alex-stiff}
that the result is very reliable as well, as long as the energies of the
states are not too far away from the true ground-state energies.

\section{Conclusion}
Results have been presented from calculations of 
a large number of ground states of 4d Ising spin glasses. They were
obtained
using a combination of cluster-exact approximation
and a genetic algorithm. Using a huge computational effort it was
ensured that true ground states have been obtained with a high probability.

The finite size behavior of the ground-state
energy and the
stiffness energy have been investigated. By performing a $L\to\infty$
extrapolation, the ground-state energy per spin for the infinite system
is estimated to be  $e_0^{\infty}=-2.095(1)$. The absolute value of the
stiffness energy increases with system size and shows a 
$\langle |\Delta(L)|\rangle\sim L^{\Theta_S}$ 
behavior with $\Theta_S=0.64(5)$. For
systems with a Gaussian distribution of the bonds qualitatively similar
results are expected, since the ordering behavior depends only on the
sign of the interactions and not on their magnitudes.

A more detailed study of the ground-state landscape of 4d systems,
similar to \cite{alex-3d}, requires more than $n_G\approx 3$ ground
states per realization to be calculated. Since this requires a
substantial higher computational effort, it remains to be done for
the future.

\section{Acknowledgements}

The author thanks A.P. Young for interesting discussions, critical
reading of the manuscript and for the
allocation  of computer time on his workstation cluster at the
University of California in Santa Cruz.
This work was suggested  by him during the ``Monbusho Meeting'' held
at the {\em Fondation Royaumont} near Paris.
The author was supported by the Graduiertenkolleg
``Modellierung und Wissenschaftliches Rechnen in 
Mathematik und Naturwissenschaften'' at the
{\em In\-ter\-diszi\-pli\-n\"a\-res Zentrum f\"ur Wissenschaftliches
  Rechnen} in Heidelberg and the
{\em Paderborn Center for Parallel Computing}
 by the allocation of computer time. Financial support was provided  
 by the DFG ({\em Deutsche Forschungsgemeinschaft}) and
the organizers of the ``Monbusho Meeting''.

\newcommand{\captionEnergy}
{Average ground-state energy $e_0$ per spin  as a function of system
  size $L$. The line shows a fit to $e_0(L)=e_0^{\infty} + a*L^{-b}$
  resulting in $e_0^{\infty} = -2.095(1)$ as estimate for the
  ground-state energy of the infinite system.
}

\newcommand{\captionPStiff}
{ Distribution of the stiffness energy $\Delta=E^a-E^p$ for system sizes
$5\times 5 \times 5 \times 5$ and $7\times 7 \times 7 \times 7$. $E^a$
and $E^p$ are the total ground-state energies for periodic and
antiperiodic boundary conditions in one direction, while for the other
three directions always periodic boundary conditions are
imposed. Lines  are guide to the eyes only.}

\newcommand{\captionStiffness}
{Average Stiffness energy $\langle |\Delta| \rangle $ as 
function of system size $L$ on log-log scale. The line
represents the function $|\Delta(L)|=aL^{\Theta_S}$ with $\Theta_S=0.65(4)$. 
The increase of $\langle |\Delta| \rangle$
with system size indicates, that for 4d Ising spin glasses an
ordered phase exists below a non-zero temperature $T_c$.}

\begin{figure}[ht]
\begin{center}
\myscalebox{\includegraphics{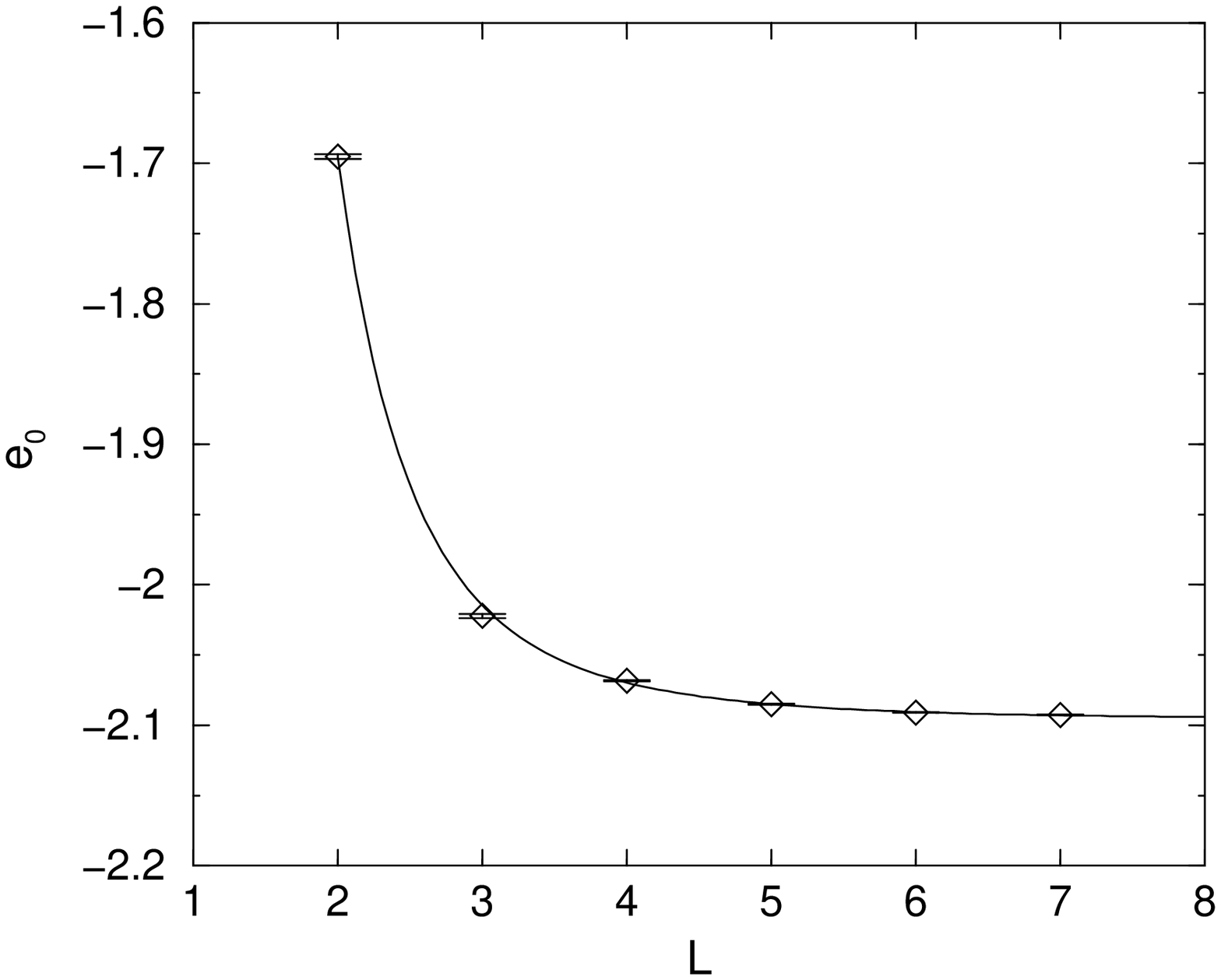}}
\end{center}
\caption{\captionEnergy}
\label{figEnergy}
\end{figure}

\begin{figure}[ht]
\begin{center}
\myscalebox{\includegraphics{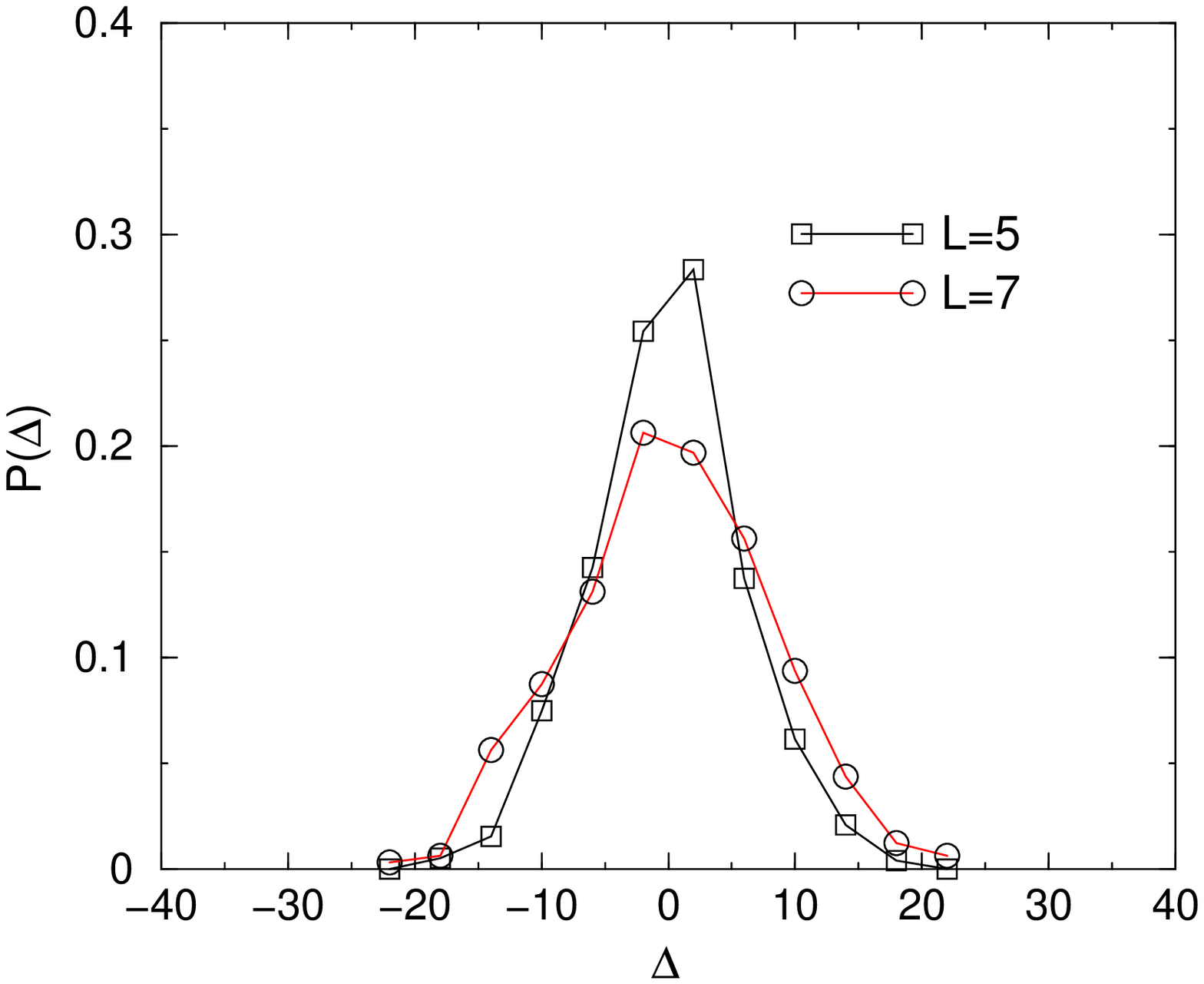}}
\end{center}
\caption{\captionPStiff}
\label{figPStiff}
\end{figure}

\begin{figure}[ht]
\begin{center}
\myscalebox{\includegraphics{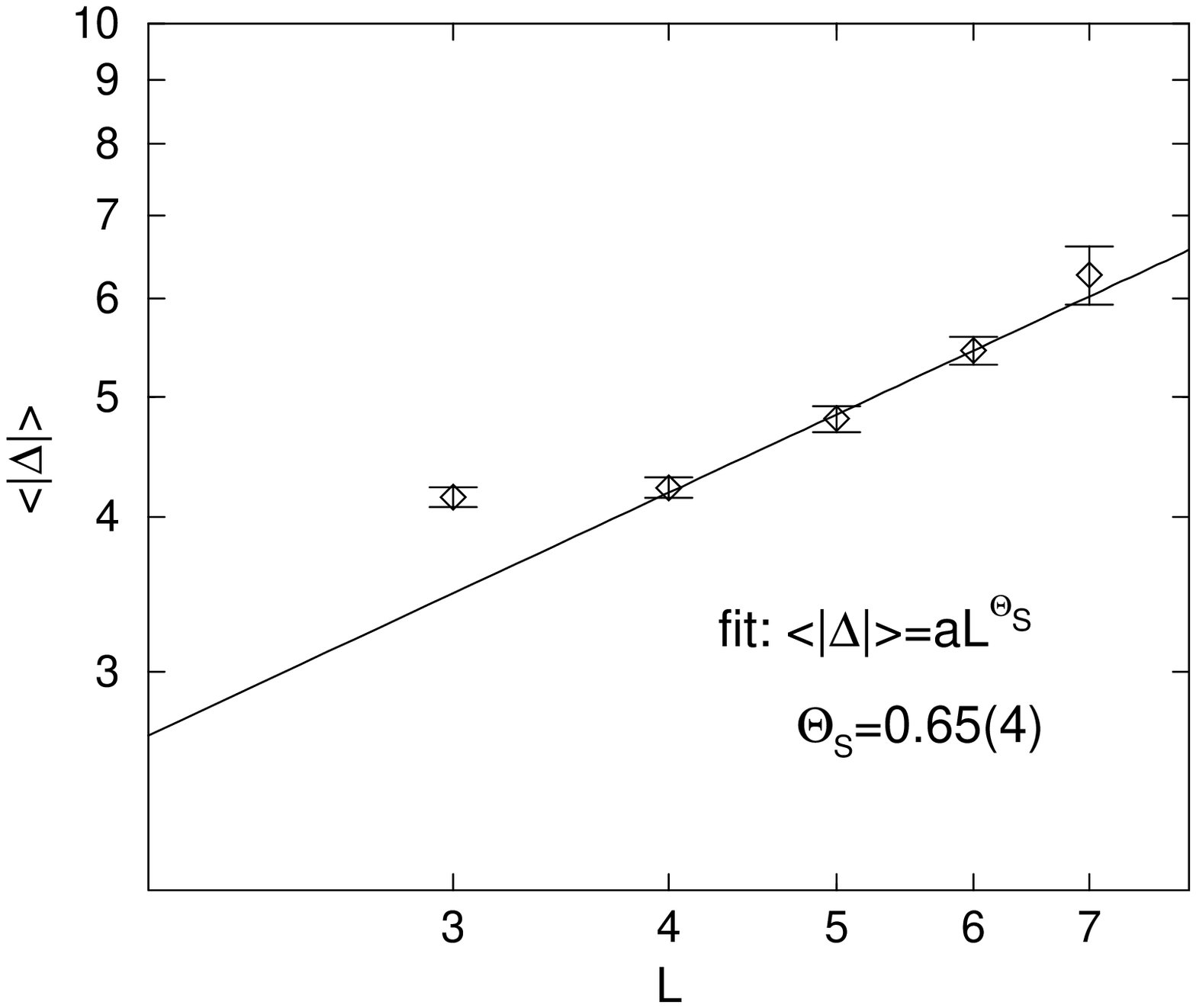}}
\end{center}
\caption{\captionStiffness}
\label{figStiffness}
\end{figure}

\begin{table}[h]
\begin{center}
\begin{tabular}{cccccc}
\hline
$L$ & $M_i$ & $\nu$ & $n_{\min}$ & $\tau$ (sec) & $N_L$ \\ \hline
2 & 16 & 1 & 1 & 0.04 & 10000 \\
3 & 16 & 4 & 4 & 3 & 9000 \\
4 & 16 & 4 & 4 & 14 & 2000 \\
5 & 256 & 6 & 10 & 4800 & 1000 \\
6 & 256 & 6 & 10 & 7300 & 1300 \\
7 & 512 & 12 & 20 & 14000 & 400 
\end{tabular}
\end{center}
\caption{Simulation parameters: $L$ = system size, $M_i$ = initial size of
population, $\nu$ = average number of offspring per configuration, $n_{\min}$
= number of CEA minimization steps per offspring, $\tau$ = typical computer
time per ground state on a 80MHz PPC601, $N_L$ = number of realizations
of the random variables.}
\label{tab_parameters}
\end{table}

\end{document}